\documentclass[11pt]{article} 

\usepackage[scale={.8,.8}]{geometry}
\usepackage[intlimits]{amsmath}
\usepackage{amssymb}
\usepackage{bbm}
\usepackage{enumerate}
\usepackage{cite} 

\newcommand{\be}{\begin{equation}}
\newcommand{\ee}{\end{equation}}
\newcommand{\bea}{\begin{eqnarray}}
\newcommand{\eea}{\end{eqnarray}}

\newcommand{\et}{\ensuremath{\tilde{e}}}
\newcommand{\at}{\ensuremath{\tilde{a}}}
\newcommand{\vt}{\ensuremath{\tilde{v}}}

\newcommand{\omegat}{\ensuremath{\widetilde{\omega}}}
\newcommand{\Mt}{\ensuremath{\widetilde{M}}}

\newcommand{\Xt}{\widetilde{X}}

\newcommand{\nut}{\widetilde{\nu}}
\newcommand{\jt}{\widetilde{j}}
\newcommand{\etat}{\tilde{\eta}}

\newcommand{\Nn}{\mathcal{N}}
\newcommand{\Pp}{\mathbb{P}}
\newcommand{\Ss}{\mathcal{S}}
\newcommand{\Ppt}{\widetilde{\mathbb{P}}}

\newcommand{\Zbb}{\mathbb{Z}}
\newcommand{\Rbb}{\mathbb{R}}
\newcommand{\e}{\mathrm{e}}

\allowdisplaybreaks

\begin{document}

\date{}
\rightline{HD-THEP-08-28}

\vskip 2cm

\pagestyle{empty}
\begin{center}
{\LARGE\bf  Type~IIB Flux Vacua from M-theory \\ \vskip 2mm  via F-theory}
\\[2.1em]

\vskip 2cm

{\Large Roberto Valandro}\\

\null

\noindent 
{\it Institut f\"ur Theoretische Physik, Universit\"at Heidelberg,\\
 Philosophenweg 16 und 19, D-69120 Heidelberg, Germany}
\\[2.1ex]
 \vfill

\end{center}

\begin{abstract}
We study in detail some aspects of duality between type~IIB and M-theory. We focus on the duality between type~IIB string theory on $K3\times T^2/\Zbb_2$ orientifold and M-theory on $K3\times K3$, in the F-theory limit. We give the explicit map between the fields and in particular between the moduli of compactification, studying their behavior under the F-theory limit. Turning on fluxes generates a potential for the moduli both in  type~IIB and in M-theory. We verify that the type~IIB analysis gives the same results of the F-theory analysis. In particular we check that the two potentials match.
\end{abstract}

\vfill 
\leftline{ r.valandro@thphys.uni-heidelberg.de}
\newpage

\pagestyle{plain}

\newpage

\section{Introduction}


One of the most studied and phenomenologically fruitfull set of string vacua is given by flux compactifications of type~IIB string theory on Calabi--Yau orientifolds. In these constructions a lot of phenomenological issues have been achieved, such as 
moduli stabilization \cite{gkp,kklt}, generation of large hierarchies by warping \cite{gkp,Dasgupta:1999ss} or by large extradimensional volume \cite{bbcq}, fine-tuning of the cosmological constant \cite{BouPol,Denef:2004ze} (for reviews see \cite{g05,dk06,bkl06})
The final goal would be to obtain global type~IIB models that describe all experimental observations. This goal is far to be achieved, even if good local constructions exist \cite{bhv08,dw08,cmq03,vw05,ms04}. 
In particular the phenomenologically promising landscape of D7-brane configurations is still not well understood, especially in presence of both bulk 3-form fluxes and worldvolume 2-form fluxes. A suitable language to describe these constructions is provided by F-theory \cite{Vafa:1996xn,Sen:1997gv} (see \cite{DenefRev} for a review). Mapping the results from F-theory to usual type~IIB theory is not in general simple. This map will be the main subject of this paper.

F-theory is a geometrical way to describe type~IIB vacua with D7-branes. In presence of D7-branes the axiodilaton $\sigma$ is generically non-constant on the compact manifold $B_6$. Because of its transformation properties under the $SL(2,\Zbb)$ symmetry of type~IIB, this complex field can be associated with the complex structure of a torus, that is fibred over the compact manifold $B_6$. This allows to encode the $B_6$ geometric data and the D7-brane data in an eight dimensional manifold that is a $T^2$-fibration over~$B_6$. 

This geometrical description of 4d type~IIB configurations can also be understood by duality with M-theory. Consider M-theory compactified on $T^2=\Ss_M^1\times \Ss_A^1$ with complex structure $\sigma$. Reducing M-theory to type~IIA on $\Ss_M^1$ and then T-dualizing along $\Ss_A^1$ we get type~IIB on $\Ss_B^1$, where the radius $R_B$ of $\Ss_B^1$ is the inverse of $R_A$. The limit in which the size of $T^2$ goes to zero corresponds to the decompactification limit in type~IIB ($R_B\rightarrow \infty$). It is called the F-theory limit. So, if we start from M-theory on an eight dimensional manifold $Y_8$, that is a $T^2$-fibration over a six dimensional manifold $B_6$, then we end up with type~IIB on $B_6$ with varying axiodilaton $\sigma$, given by the complex structure of the fibre. The deformations of $Y_8$ include the geometric moduli of $B_6$, the axiodilaton and the motion and recombination of the D7-branes.

The duality between type~IIB and M-theory has been extensively used to study type~IIB flux compactifications (see e.g. \cite{Dasgupta:1999ss,Denef:2005mm,Denef:2004ze}). For example, instead of considering the 3-form flux generated superpotential, one can consider the M-theory superpotential generated by 4-form flux on $Y_8$ \cite{GVW}. These methods allowed, for instance, to realize that the D7-brane moduli are fixed by 3-form fluxes. Actually, in many cases it is simpler to work using the M-theory language than explicitely in type~IIB. This is due to the fact that type~IIB objects with different nature are described in a unified way in M-theory. For example, as we said, both type~IIB geometric moduli and D7-brane moduli are mapped to geometric moduli in M-theory. Moreover 3-form fluxes and 2-form fluxes are all encoded into 4-form M-theory fluxes. Deriving results in M-theory is then more immediate when we want to consider these objects. The non-trivial step can be to map these results to type~IIB and to to take the F-theory limit appropriately. It is then important to take confidence with the duality map and see how the different type~IIB fields are described in M-theory, before and after the F-theory limit. 
This is the scope of this work. We will explain in detail how the duality works in a particular compactification, giving a useful map between the fields on the two sides. We will consider one largely studied type~IIB compactification, {\it i.e.} type~IIB on $K3\times T^2/\Zbb_2$ orientifold (with D7-branes) \cite{TripTriv,Fer,Fer2}, as well as its dual M-theory compactification on $K3\times K3$  \cite{Dasgupta:1999ss,Lust,Kachru,BergMay,AspKall,LustKall,BHLV}. In this case we are able to find a reliable and efficient dictionary between the two sides of the duality. The procedure detailed here can in principle be used in more complicated cases.

We are in particular interested in studying backgrounds with fluxes. We will consider both 3-form bulk fluxes and 2-form fluxes on the D7-brane worldvolume. As an application of the duality map that we describe in this paper, we show how the type~IIB potential generated by these fluxes can be obtained by taking the M-theory potential, applying the map and doing the F-theory limit. In type~IIB on CY orientifolds the system with 3-form fluxes and D7-branes with 2-form fluxes is in general not deeply understood yet. In M-theory, this system is mapped to geometric background with 4-form fluxes, that are more easy to control. For this reason, we believe that it is useful to do the check on the potentials and, in particular, to explain the M-theory origin of different contributions of the type~IIB scalar potential. Starting from the M-theory 4-form flux potential and applying the duality and the limit, we will obtain precisely the type~IIB flux potential found by \cite{Fer} in the context of gauged supergravity. This procedure can in principle be used in more complicated cases, where it is simple to compute the M-theory potential, but difficult to derive 2-form and 3-form combined flux potential in type~IIB. Other times the potentials are easily derived in both theories, but the minimization is easier on the M-theory side. In this case the map is useful to translate the results to type~IIB. For example in the particular compactification we have studied, the M-theory language is more usefull to study type~IIB moduli stabilization \cite{BHT,BHLV}. 

\

Let us summarize the structure of the paper. In Section \ref{IIBsec}, we start with a review of type~IIB on $K3\times T^2/\Zbb_2$ (as treated in \cite{Fer,Fer2}), in which we focus on the features that we want to derive by duality from M-theory. In particular we give the form of the flux potential.

As we have explained above, we have a clean duality between M-theory on $K3\times K3$ and type~IIB on $K3\times T^2/\Zbb_2\times \Ss_B^1$. To obtain the 4d type~IIB compactification we have to send the radius of $\Ss_B^1$ to infinity. On the M-theory side we have to take the limit of zero fibre size (F-theory limit). For this reason, in Section \ref{Sec:duality} we study the map between these two backgrounds, before the limit. At first we compactify type~IIB on $K3\times T^2/\Zbb_2\times \Ss_B^1$. We list all the 3d fields coming from this compactification. Then we compactify M-theory on $K3\times K3$ and relate all the resulting fields to the type~IIB ones. We divide the 3d spectrum into a set of scalars and a set of vectors and we give the map for both sets. In 3d a vector is dual to a scalar, so this separation could appear arbitrary. The splitting becomes clear after the F-theory limit. The effect of this limit on the 3d type~IIB fields is simple to derive. Applying the duality, one can understand the behavior of the M-theory fields under it. In particular, we see that what we have chosen as 3d vectors combine with 3d scalars to form 4d vectors. These 3d fields
have a slightly different nature in M-theory: the vectors come from reduction of the three-form $C_3$, while the scalars are metric deformations. This fact can be used to guess the effects of fluxes on vectors, just looking at the M-theory potential for the geometric moduli. In fact, we see that switching on some M-theory fluxes lift the 3d geometric moduli related to the 4d vectors; from this one could guess that the corresponding vectors get a mass from fluxes. This is precisely what happens: it has been seen both in type~IIB \cite{Fer,Fer2} and directly in M-theory \cite{BHLV}.

In Section \ref{Sec:potential} we apply the duality map to the flux potential. Thanks to the analysis of the previous section, we are able to map the 3d M-theory flux potential to the 4d type~IIB flux potential, both in the situation with only 3-form fluxes turned on and when also $F_2$ D7 fluxes are switched on. The M-theory full scalar potential has been written down explicitely and studied in \cite{BHLV}. Applying the duality map to this potential we find the type~IIB scalar potential studied in \cite{Fer,Fer2}. We also find that the supersymmetry conditions are the same.

We conclude with some Appendices. In Appendix \ref{DualWithHet} we describe the Heterotic dual of the set of vacua analyzed in this paper. Heterotic theory $E_8\times E_8$ on $T^3$ is dual to M-theory on $K3$ \cite{Witten:1995ex}. This leads to the duality between type~IIB on $K3\times T^2/\Zbb_2$, Heterotic on $K3\times T^2$, and M-theory on $K3\times K3$ in the F-theory limit. Using this duality we will give the map between the fields described in the paper, and the Heterotic ones. This could be useful to study flux backgrounds in Heterotic theory.

\vskip 15mm

\section{Type~IIB on \boldmath$K3\times T^2/\Zbb_2$}\label{IIBsec}

In absence of fluxes, type~IIB compactified on the $K3\times T^2/\Zbb_2$ orientifold gives ungauged 4d $\Nn=2$ supergravity with a certain content of hypermultiplets and vector multiplets \cite{deWit:1984px,Andrianopoli:1996cm,Andrianopoli:1996vr}. The introduction of 3-form fluxes gauges some isometries of the moduli space by some vectors at disposal. The supergravity analysis of these flux vacua (found by \cite{TripTriv}) is presented in \cite{Fer}. In this section we will briefely review their results.

\

Before the orientifold projection, type~IIB on $K3\times T^2$ has $\Nn=4$ supersymmetries in 4d. The orientifold action is give by $(-1)^{F_L}\Omega_p \Zbb_2$, where the $\Zbb_2$ inverts the two coordinates of $T^2$. This introduces orientifold 7-planes, wrapped on $K3$ and situated at the $\Zbb_2$ singularities of $T^2/\Zbb_2$.

After orientifolding, the 4d spectrum is \cite{TripTriv,Fer} (we write only the bosonic fields):
\begin{description}
\item 1 gravity multiplet: $(g_{\mu\nu},A^0_\mu)$;
\item 3 vector multiplets: $(A_\mu^i,\Phi^i)$ with $i=1,2,3$ and $\Phi^i$ complex scalars;
\item 20 hypermultiplets constructed using the 80 scalars $e_i^m$ ($m=1,...,19$, $i=1,2,3$), $C^I$ ($I=1,...22$) \\ and $\phi$.
\end{description}

Let us see how this spectrum comes from compactification. The 4d metric is obviously derived by reducing the 4d part of the 10d metric. The four vectors $A_\mu^K$ ($K=0,...,3$) come from the KK expansion of the type~IIB 2-forms $B_2$ and $C_2$. In fact these fields are odd under $(-1)^{F_L}\Omega_p$, and so they must me expanded into forms odd under $\Zbb_2$. The vectors are the result of expanding $B_2$ and $C_2$ on the two odd 1-cycles of $T^2$.

The scalars come from various type~IIB fields. 
\begin{itemize}
\item The three complex scalars in the vector multiplets are denoted by 
\be
 \rho = \rho_1+i\rho_2\:, \qquad\qquad \tau = \tau_1+i\tau_2 \:, \qquad\qquad \sigma=C_0+ie^{-\varphi_0} \:,\nonumber
\ee
where $\rho_1$ comes from the $C_4$ field expanded on the volume form of $K3$, $\rho_2$ is the volume modulus of $K3$, $\tau$ is the complex structure modulus of $T^2$ and $\sigma$ is the type~IIB axio-dilaton.
\item The $C^I$ ($I=1,...,22$) scalars in the hypermultiplets come from $C_4$ expanded on 4-forms $\eta_I\wedge Vol_T$, where $\{\eta_I\}$ is a basis of $H^2(K3)$ and $Vol_T$ is the volume form of $T^2/\Zbb_2$.
\item The scalar $\phi$ is the volume modulus of $T^2/\Zbb_2$.
\item The 57 scalars $e^b_i$ ($b=1,...,19$ and $i=1,2,3$), are the metric moduli of $K3$ that control its hyperK\"ahler structure (See Appendix \ref{AppendixK3}).
\end{itemize}

The scalars \label{scalarspg} listed above are the moduli of this specific compactification \cite{Fer}. 

\

After orientifolding type~IIB on $K3\times T^2$, we are left with four O7-planes, one at each singularity of $T^2/\Zbb_2$. They are wrapped on $K3$ and span the spacetime directions. This introduces a D7-charge on $T^2/\Zbb_2$, that must be cancelled. This is done by introducing 16 D7-branes wrapped on $\Rbb^{1,3}\times K3$. The D7-branes introduce new fields, since on the worldvolume of each D7-brane, there lives an 8d SYM theory. In 4d, this gives: 
\begin{description}
\item 16 vector multiplets: $(A_\mu^\vartheta,z^\vartheta)$ with $\vartheta=1,...,16$, where $z^\vartheta$ are the scalars that parametrize the positions of the D7-branes on $T^2/\Zbb_2$.
\end{description}
When some of the branes are on top of each others, there is an enhancement of the gauge group and new massless vector multiplets arise. A special case is when 4 D7-branes are placed on top of each orientifold plane: then the D7-charge is cancelled locally and the gauge group is enhanced from $U(1)^{16}$ to $SO(8)^4$.

D7-branes and O7-planes wrapped on a curved manifold give a negative contribution to the D3-charge. In particular, one D7-brane wrapped on $K3$ gives contribution $-1$ to the D3-charge, while an O7-plane gives $-2$. Hence the total D3-charge of the 16 D7's and the 4 O7's is $-24$. It can be cancelled by introducing D3 branes or by turning on fluxes. In fact, the tadpole cancellation condition is\footnote{We work in unit where the quantized fluxes have integral coefficients with respect to integral bases.}
\begin{equation}
N_{flux}^{OR} + N_{D3} = 24\:, \qquad\qquad \mbox{where }\qquad N_{flux}^{OR} = \int_{K3\times T^2/\Zbb_2}H_3\wedge F_3 \:.
\end{equation}
$N_{D3}$ is the number of D3-branes and $N_{flux}^{OR}$ is the D3-charge carried by the fluxes.
In what follows, we will take $N_{D3}=0$, as D3 branes would introduce new fields.

\subsection{Fluxes and Gauging}

Turning on 3-form fluxes on $K3\times T^2/\Zbb_2$ gauges some isometries of the quaternionic manifold, by the four vectors in the hypermultiplets. 

The 3-form flux can be expanded on a basis of harmonic 3-forms of $K3\times T^2/\Zbb_2$. We will consider the basis
$\{\eta^I\wedge dx,\eta^I\wedge dy\}$ ($\{\eta^I\}$ is a basis of $H^2(K3)$ and $(x,y)$ are the flat coordinates on $T^2$). $H^2(K3)$ is isomorphic to $\Rbb^{3,19}$ with the inner product given by the wedge product (see \eqref{K3modmetric}). We will split the index $I=1,...,22$ into $i=1,2,3$ and $b=1,...,19$, corresponding to taking the basis $\eta^I$ with three positive norm vectors $\eta^a$ and nineteen negative norm vectors $\eta^m$. We take this basis to be orthonormal and with the vectors parallel to integral forms.

The expansion of the 3-form fluxes $F_3$ and $H_3$ on this basis are \cite{TripTriv}
\begin{eqnarray}\label{3flux}
 F_3 &=& \frac{1}{\sqrt{2}}\left\{(f^i_0 - f^i_2)\,\eta^i\wedge dx + (f^i_1-f^i_3) \,\eta^i\wedge dy +(h^b_0-h^b_2)\,\eta^b\wedge dx +(h^b_1-h^b_3)\,\eta^b\wedge dy \right\}\:,\nonumber\\ \\
 H_3 &=&  \frac{1}{\sqrt{2}}\left\{(f^i_1+f^i_3)\,\eta^i\wedge dx - (f^i_0+f^i_2)\,\eta^i\wedge dy +(h^b_1+h^b_3)\,\eta^b\wedge dx -(h^b_0+h^b_2)\,\eta^b\wedge dy  \right\}\:.\nonumber
\end{eqnarray}
The coefficients are constrained by the requirement that $F_3$ and $H_3$  be integral forms.
The fluxes \eqref{3flux} have the following charge\footnote{We are using the normalization $\int_{T^2/\Zbb_2}dx\,dy = 1$.}:
\begin{equation}\label{F3H3charge}
 N_{flux}^{OR} = \int_{K3\times T^2/\Zbb_2}H_3\wedge F_3 = \frac12 \left(f_0^2-f_2^2+f_1^2-f_3^2-h_0^2+h_2^2-h_1^2+h_3^2\right)\:.
\end{equation}

\

The isometries that are gauged are the shift symmetries related to the axions $C^I$:
\begin{equation}
 D_\mu C^b = \partial_\mu C^b+h^b_K A_\mu^K \qquad\qquad D_\mu C^i = \partial_\mu C^i+f^i_K A_\mu^K
\end{equation}
with $b=1,...,19$, $i=1,2,3$ and $K=0,...,3$. $h^b_K$ and $f^i_K$ are the coupling constants related to the 3-form fluxes \eqref{3flux} \cite{Fer}. When performing the dimensional reduction, the kinetic terms for the axions come with these covariant derivatives.

\

Different choices of the coupling constants give different kinds of vacua:
\begin{enumerate}
\item When $f^i_K=0$ $\forall i,K$ and $h^b_K=0$ $\forall b,K$ except $h_2^1\equiv \ell_1$ and $h_3^2\equiv \ell_2$, then the corresponding configurations have $\Nn=2$ supersymmetries. The vectors that take mass, because of gauging, are the vector partner of $\tau$ and $\sigma$. 
\item \label{caso2} When $h^b_K=0$ $\forall b,K$ and $f^i_K=0$ $\forall i,K$ except $f_0^1\equiv g_0$ and $f_1^2\equiv g_1$, then we have $\Nn=0,1$ configurations. In particular we have $\Nn=1$ when $g_0=g_1$. The vectors that acquire mass are the graviphoton and the partners of the $K3$ volume. 
\item When all $\ell_1,\ell_2,g_0$ and $g_1$ are different from zero, then the configuration is still $\Nn=0$ ($g_0\not = g_1$) or $\Nn=1$ ($g_0 = g_1$), but in this case all the four vectors get mass. 
\end{enumerate}

\subsection{Scalar Potential}

In this section we will review the analysis of the scalar potential in the configurations (2) described above, {\it i.e.} $\ell_1=\ell_2=0$ and $g_0,g_1$ different from zero. The full treatment is presented in \cite{Fer}. Here we will only report the results.

The potential for the scalar fields can be computed, for abelian gauging, following \cite{Andrianopoli:1996cm}. Once we express it in terms of the scalars introduced at page \pageref{scalarspg} and we take $h^b_K=0$ $\forall b,K$, we get
\begin{eqnarray}\label{PotPrimo}
 V &=& e^{2\phi}\,e^{\widetilde{\mathcal{K}}}\left\{4\,e^{\widehat{\mathcal{K}}}\left(\delta_{ij}+2e_i^b e_j^b\right) f_K^i f_H^j \, X^K \bar{X}^H - 2\left(\delta_{ij}+e_i^b e_j^b\right)f_K^i f_H^j \eta^{KH} \right\}\:,
\end{eqnarray}
where $\eta^{KH}=$diag$(+1,+1,-1,-1)$, $\widetilde{\mathcal{K}},\widehat{\mathcal{K}}$ are defined by
\begin{equation}\label{KpotIIB}
 \widetilde{\mathcal{K}} = -\log\,i(\rho-\bar{\rho})  \qquad\qquad \widehat{\mathcal{K}} = -\log\,\frac{1}{2}i(\tau-\bar{\tau})i(\sigma-\bar{\sigma})\:,
\end{equation}
and the $X^K$'s are functions of $\tau$ and $\sigma$:
\begin{equation}\label{X0123}
 X^0=\frac12 (1-\tau\sigma) \qquad X^1=-\frac12 (\tau+\sigma) \qquad X^2=-\frac12 (1+\tau\sigma) \qquad X^3=\frac12 (\tau-\sigma)\:.
\end{equation}
Taking $f_0^1\equiv g_0$ and $f_1^2\equiv g_1$ and the others equal to zero, the potential \eqref{PotPrimo} becomes
\begin{eqnarray}\label{potIIB1}
 V &=& e^{2\phi}\,e^{\widetilde{\mathcal{K}}}\left\{4\,e^{\widehat{\mathcal{K}}}\left[g_0^2 |X^0|^2 (1+ 2e_1^be_1^b) + g_1^2 |X^1|^2 (1+ 2e_2^be_2^b) + \right.\right.\\
 && \left.\left. + 2 g_0 g_1 e_1^be_2^b (X^0\bar{X}^1+\bar{X}^0X^1)\right] -2 \left[ g_0^2  (1+ e_1^be_1^b) + g_1^2 (1+ e_2^be_2^b)\right]\right\}\:.\nonumber
\end{eqnarray}

This potential has been proved to be positive definite and to take minima at $V=0$ \cite{Fer}. This condition is fullfilled by
\begin{equation}
 \tau=\sigma=i \qquad \mbox{and} \qquad e_1^b = e_2^b = 0 \:.
\end{equation}

The extremum condition does not fix the scalars $\phi$, $\rho$, $e_3^b$ and the remaining $C^I$ ($C^b$ with $b=1,2$ disappear from the spectrum because of gauging). All these scalars remain massless. 

If $g_0=g_1$ the vacua preserve $\Nn=1$ supersymmetry. The massless scalars coming from $\phi$, $\rho$, $e_3^b$ and the remaining $C^I$ organize in massless chiral multiplets. If we change the fluxes such that $g_0\not =g_1$, the vacua do not preserve supersymmetry anymore.

When we turn on also non-zero $\ell_1$ and $\ell_2$, we get a potential also for the scalars $e_3^1,e_3^2$ and a mass for all the four vectors
\cite{Fer}\label{ell1ell2IIBresults}. If moreover one takes into account also the D7 moduli, the potential, at the extremum of the $e_i^b$ scalars, has the following form \cite{Fer2}:
\begin{equation}
 V =  e^{2\phi} e^{\widetilde{\mathcal{K}}}
\left\{4e^{\widehat{\mathcal{K}}}\left[g_0^2\left|X^0\right|^2+g_1^2\left|X^1\right|^2+\ell_1^2\left|X^2\right|^2+\ell_2^2\left|X^3\right|^2
\right] -2\left(g_0^2+g_1^2\right)  \right\}\:,
\end{equation}
where the expressions for $\widehat{\mathcal{K}}$ and $X^K$ have been changed to
\begin{equation}\label{KpotIIBbis}
 \widehat{\mathcal{K}} =
-\log\,\left[\frac{1}{2}\left(i(\tau-\bar{\tau})i(\sigma-\bar{\sigma})-\sum_{\vartheta=1}^{16}\frac{(z^\vartheta-\bar{z}^\vartheta)^2}{2}\right)\right]
= -\log\,\left[2\left(\tau_2\sigma_2
-\sum_{\vartheta=1}^{16}\frac{(y^\vartheta)^2}{2}\right)\right]
\end{equation}
\begin{equation}\label{X0123bis}
 X^0=\frac12 \left(1-\tau\sigma+\frac{z^2}{2}\right) \,\,\,\,
X^1=-\frac12 (\tau+\sigma) \,\,\,\, X^2=-\frac12 \left(1+\tau\sigma -
\frac{z^2}{2}\right) \,\,\,\, X^3=\frac12 (\tau-\sigma)\:.\nonumber
\end{equation}
Here $z^2=\sum_\vartheta (z^\vartheta)^2$ and $z^\vartheta=x^\vartheta + i y^\vartheta$ are the positions of the 16 D7-branes on $T^2/\Zbb_2$.

\

If we now gauge the remaining isometries by using the gauge fields on the
worldvolume of the D7-branes, the potential (at $e_i^b=0$) gets a new contribution and becomes
\begin{equation}\label{potentialIIBallIsom}
 V =  e^{2\phi} e^{\widetilde{\mathcal{K}}}
\left\{4e^{\widehat{\mathcal{K}}}\left[g_0^2\left|X^0\right|^2+g_1^2\left|X^1\right|^2+\ell_1^2\left|X^2\right|^2+\ell_2^2\left|X^3\right|^2
        + \sum_{\vartheta=1}^{16} \ell_{\vartheta+2}^2 \left|X^{\vartheta+3}\right|^2\right]
-2\left(g_0^2+g_1^2\right)\right\}\:,
\end{equation}
where $X^{\vartheta+3}=\frac{z^\vartheta}{\sqrt{2}}$.

\vskip 8mm
\section{Type~IIB on \boldmath$K3\times T^2/\Zbb_2\times \Ss^1$ and its M-Theory Dual}\label{Sec:duality}

Type~IIB on $K3\times T^2/\Zbb_2$ can be seen as the limit of type~IIB on $K3\times T^2/\Zbb_2\times \Ss_B^1$ when the radius of $\Ss^1_B$ goes to infinity. This 3d compactification turns out to be dual to M-theory on $K3\times K3$.

In the next section we will study the 3d type~IIB spectrum and we will recover the 4d spectrum by taking the limit $R_B\rightarrow \infty$. Then we will describe M-theory on $K3\times K3$ and we will see what is the dual limit that should give 4d spectrum.

\subsection{Type~IIB on \boldmath$K3\times T^2/\Zbb_2\times \Ss^1$}

Let us take type~IIB on $K3\times T^2/\Zbb_2\times \Ss^1_B$ and consider the resulting 3d spectrum.

We start by considering what are the 3d $U(1)$ vectors (that we will denote by a hat to distinguish them form the 4d ones)\footnote{In this list we did not include the 22 vectors coming from $C_4$ on 2-cycles of $K3$ and on $\Ss^1_B$. In fact we will count them among the scalars, as $C_4$ satisfies a self-duality condition that identifies the 22 vectors with the 22 scalars (we remember that in 3d the vectors are dual to scalars).
}:
\begin{description}
\item 1 vector $\hat{g}_\mu$ from the metric $g$ with one index on $\Ss^1_B$.
\item 4 vectors $\hat{A}_\mu^K$ ($K=0,...,3$) from $B_2,C_2$ with one index on $T^2/\Zbb_2$.
\item 1 vector $\hat{C}_{4\mu}$ from $C_4$ with two index on $T^2/\Zbb_2$ and one on $\Ss^1_B$.
\item 16 vectors $\hat{A}_\mu^\vartheta$ ($\vartheta=1,...,16$) from the 16 D7-branes wrapped on $\Rbb^{1,2}\times K3\times \Ss^1_B$.
\end{description}

\

Let us now consider the (real) scalars (again, we will denote the 3d spectrum with a hat):
\begin{description}
\item 58 scalars $\hat{\rho}_2,\hat{e}_i^b$ from the metric on $K3$.
\item 22 scalars $\hat{C}^I$ $I=1,...,22$ from $C_4$ with two indeces on $T^2/\Zbb_2$ and two on a 2-cycle of $K3$.
\item 3 scalars $\hat{\phi},\hat{\tau}$ from the metric on $T^2/\Zbb_2$.
\item 1 scalar $\hat{r}_B$ from the metric on $\Ss^1_B$.
\item 2 scalar $\hat{\sigma}$ from the axio-dilaton.
\item 4 scalars $\hat{A}_B^K$ ($K=0,...,3$) from $B_2,C_2$ with one index on $T^2/\Zbb_2$ and one on $\Ss^1_B$.
\item 32 scalars $\hat{x}^\vartheta,\hat{y}^\vartheta$ ($\vartheta=0,...,16$) from the positions of the 16 D7-branes on $T^2/\Zbb_2$.
\item 16 scalars $\hat{A}_B^\vartheta$ ($\vartheta=0,...,16$) from the D7-brane gauge fields along $\Ss^1_B$.
\end{description}

Summarizing, we have 58+22+58 = 138 scalars and 22 vectors.

\

\label{IIBlimit}
Let us see what happens if we let the $\Ss^1_B$ radius go to infinity. In this case we recover type~IIB on $K3\times T^2/\Zbb_2$. The Kaluza-Klein modes relative to compactification on $\Ss^1_B$ become massless, giving the fields the dependence on the fourth coordinate. The scalars $\hat{A}_B^K$ become the fourth component of the vectors $\hat{A}_\mu^K$, resulting in the 4d vectors $A_\mu^K$. In the same way $\hat{A}_\mu^\vartheta$ and $\hat{A}_B^\vartheta$ combine to give the 4d vectors $A_\mu^\vartheta$ on the D7-branes worldvolume. The vector $\hat{g}_\mu$ and the scalar $\hat{r}_B$ combine to give the 4d metric fluctuations $g_{\mu\nu}$. The vector $C_{4\mu}$ becomes a 4d 2-form that dualizes to the 4d scalar $\rho_1$. We are left with the 80 real 4d scalars $C^I,e_i^b,\phi$, with the 3 complex scalars $\rho,\tau,\sigma$ and with the 16 complex scalars associated with the D7 positions. We have recovered the spectrum of the section \ref{IIBsec}. In particular the Wilson lines disappear from the moduli space, as they become pure gauge (while the relative propagating degrees of freedom become the fourth component of 4d vectors). This corresponds to the fact that the limit changes the topology of the space (from $\Rbb^{2,1}\times \Ss^1$ to $\Rbb^{3,1}$).

\subsection{M-Theory on \boldmath$K3\times K3$}

M-theory is described at low energy by 11d supergravity. The bosonic fields are the metric and a 3-form $C_3$. 

We compactify M-theory on the 8d manifold $K3\times \widetilde{K3}$. The resulting 3d spectrum is\footnote{
Since in 3d a vector is dual to a scalar, the separation of the spectrum in vectors and scalars could appear arbitrary. In this case, the choice is adapted to the duality map we want to describe. The vectors are those fields that (after the F-theory limit) will become type IIB 4d vectors.}:
\begin{description}
\item 58 scalars $e_i^b$ ($i=1,2,3$ and $b=1,...,19$), describing the hyperK\"aher structure of $K3$, and the volume modulus $\nu$.
\item 22 scalars $C_3^I$ ($I=1,...,22$) from dualizing the 22 vectors coming from $C_3$ on 2-cycles of $K3$.
\item 58 scalars $\tilde{e}_j^c$ ($j=1,2,3$ and $c=1,...,19$), describing the hyperK\"aher structure of $\widetilde{K3}$, and the volume modulus $\tilde{\nu}$.
\item 22 vectors $\tilde{C}_\mu^\Lambda$ ($\Lambda=1,...,22$) from $C_3$ on 2-cycles of $\widetilde{K3}$.
\end{description}
Again we have 58+22+58 = 138 scalars and 22 vectors, like in type~IIB on $K3\times T^2/\Zbb_2\times \Ss^1_B$. In the next section we will explicitly map the fields of the two sets.

\

The curvature of $K3\times K3$ induces a negative M2-charge given by $\frac{\chi(K3\times K3)}{24}=24$ \cite{BandB}. This can be cancelled by introducing M2 branes or by fluxes. The M2-brane charge we find in M-theory is the same as the D7/07-generated D3-charge in type~IIB.

\subsection{M-Theory / Type~IIB Duality}

M-theory on a torus is dual to type~IIB on a circle \cite{Witten:1995ex}. In fact, M-theory on $\Ss^1_M\times\Ss^1_A$ is dual to type~IIA on $\Ss^1_A$ and type~IIA on $\Ss^1_A$ is T-dual to type~IIB on $\Ss^1_B$, where the radius $R_B$ is the inverse of the radius $R_A$. This can be extended to M-theory compactified on $T^2$ fibrations when the type~IIB dilaton is not constant (see also \cite{DenefRev}).

Consider type~IIB on $\mathcal{M}\times \Ss^1_B$ (with D7-branes wrapping $\Ss^1_B$). This turns out to be dual to M-theory on a $T^2$ fibration over $\mathcal{M}$, where $T^2=\Ss^1_M\times\Ss^1_A$. Let us summarize what happens to the type~IIB fields under the two dualities:
\begin{enumerate}
\item \label{pt1dual}The metric $g$ along $\mathcal{M}$ remains the metric on $\mathcal{M}$.
\item \label{pt2dual}The metric $g$ with one index along $\Ss^1_B$ becomes IIA $B_2$ with one leg on $\Ss^1_A$. It goes to $C_3$ with two indices along $T^2$.
\item \label{pt3dual}$B_2$ with no index along $\Ss^1_B$ becomes IIA $B_2$ along $\mathcal{M}$. It goes to $C_3$ with one index along $T^2$.
\item \label{pt4dual}$B_2$ with one index along $\Ss^1_B$ becomes IIA metric with one index on $\Ss^1_A$. It goes to M-theory metric elements with one index on $T^2$ and one on $\mathcal{M}$.
\item \label{pt5dual}$C_2$ with no index along $\Ss^1_B$ becomes IIA $C_3$ with one index along $\Ss^1_A$. It goes to M-theory $C_3$ with one index along $T^2$.
\item \label{pt6dual}$C_2$ with one index along $\Ss^1_B$ becomes IIA $C_1$ on $\mathcal{M}$. It goes to M-theory metric elements with one index on $T^2$ and one on $\mathcal{M}$.
\item \label{pt7dual}$C_4$ with one index on $\Ss^1_B$ becomes IIA $C_3$ on $\mathcal{M}$. It goes to M-theory $C_3$ with no index along $T^2$. (Since $C_4$ has a selfdual field strength, $C_4$ with no index on $\Ss^1_B$ goes to the same M-theory field).
\item \label{pt8dual}$C_0$ becomes IIA $C_1$ on $\Ss^1_A$, that goes to M-theory metric elements on $T^2$.
\item \label{pt9dual}The dilaton $\phi$ becomes a combination of IIA dilaton and the size of $\Ss^1_A$. It goes to M-theory metric elements on $T^2$. 
\item \label{pt10dual} The positions of the D7-branes on $\mathcal{M}$ become IIA positions of the D6 branes on $\mathcal{M}$ that go to M-theory metric elements describing the fibration of $T^2$ on $\mathcal{M}$.
\item \label{pt11dual} The $U(1)$ $A_\mu$ on the D7-branes, with no index along $\Ss^1_B$, become IIA $A_\mu$ on the D6 branes. These go to $U(1)$ coming from $C_3$ expanded along 2-forms with one index on the base and one on the fibre.
\item \label{pt12dual} The $U(1)$ $A_\mu$ on the D7-branes, with the index along $\Ss^1_B$, become IIA D6 brane positions on $\Ss^1_A$, that go to M-theory metric elements (in particular they determine the points of $T^2$ where the M-theory cycle degenerates).
\end{enumerate}

Let us explain more explicitely the last three points. The relation between type~IIB and type~IIA is the usual T-duality on one $\Ss^1$. Under that a D7-brane wrapping the $\Ss^1$ goes to a D6-brane localized on $\Ss^1$; the gauge fields living on the D7-branes go to the gauge fields living on D6-branes and on scalars describing the positions of D6-branes along the $\Ss^1$. Under the duality between type~IIA on a manifold $\mathcal{N}$ and M-theory on an $\Ss^1_M$ fibration over $\mathcal{N}$ \cite{Witten:1995ex,Townsend:1995kk} a D6-brane becomes a 7d submanifold of $\mathcal{N}$ over which the circle $\Ss^1_M$ degenerates. So the moduli describing the positions of the D6-branes in the transverse directions become, in M-theory, the metric moduli describing the fibration. Moreover, when we have two (or more) D6-branes on top of each others, the fibration develops a singularity that produces an enhancement of the gauge group.%
\footnote{The enhancement of gauge group by singularities has been extensively used to get non-abelian gauge group in M-theory compactifications on $G_2$ holonomy manifolds \cite{a00,ag04}.}

Consider now the the duality between type~IIB on $\mathcal{M}$ and M-theory on a $T^2$ fibration over $\mathcal{M}$. The D7-branes are wrapped on $\Ss_B^1$ and on a 7d submanifold of $\mathcal{M}$. The dual D6-branes are localized on $\Ss^1_A$ and span the 7d submanifold. The positions of the D6-branes in $\mathcal{M}$ are given by the positions of the D7-branes in $\mathcal{M}$, while the positions on $\Ss^1_A$ are given by the abelian Wilson lines of the D7 gauge fields along $\Ss^1_B$. Then both the moduli describing the D7 positions and the Wilson lines go to metric moduli in M-theory. If two D7-branes are on top of each other, the fibration is singular and the gauge group is enhanced to $U(2)$. To break this group to $U(1)\times U(1)$, one can either separate the two D7-branes or switch on an appropriate Wilson line on $\Ss^1_B$; in IIA both of these choices correspond to separating the D6-branes; in M-theory this is realized by metric deformations that correspond to blowing up some cycles, resolving the singularity.

\subsubsection*{Explicit map of the fields in the case $\mathcal{M}=\Rbb^{2,1}\times K3\times T^2/\Zbb_2$}\label{ExplicitMap}

We now apply the recipe given before to our case $\mathcal{M}=\Rbb^{2,1}\times K3\times T^2/\Zbb_2$. We will be able to map the 3d type~IIB fields to M-theory ones.

First, we note that the M-theory $T^2$ is fibred only over $T^2/\Zbb_2$. 
The $T^2$ fibration over $T^2/\Zbb_2$ is a $K3$, that we will call $\widetilde{K3}$. Hence type~IIB on $K3\times T^2/\Zbb_2\times \Ss^1$ orientifold is dual to M-theory on $K3\times \widetilde{K3}$, where $\widetilde{K3}$ is an elliptic fibration over $\mathbb{C}\Pp^1$. 

Requiring $\widetilde{K3}$ to admit an elliptic fibration means that there must exist at least two algebraic curves embedded in $\widetilde{K3}$: the fibre $T^2$ and the base $\mathbb{C}\Pp^1$. These are two 2-cycles $F$ and $B$ with intersection matrix
\begin{equation}
\left(\begin{array}{rr}
0 & 1\\ 1 & -2\\
\end{array}\right)\:.
\end{equation}
The Poincar\'e dual 2-forms, that we will still denote $F$ and $B$, must be orthogonal\footnote{
We have introduced the natural metric on $H^2(\widetilde{K3})$ given by the wedge product (see Appendix \ref{AppendixK3}): $(\vt \cdot \tilde{w})=\int_{\widetilde{K3}} \vt\wedge \tilde{w}$.
} to the complex structure $\omegat$ of $\widetilde{K3}$. $\widetilde{K3}$ is an hyperK\"ahler manifold, whose structure and metric are defined by a 3d~positive norm subspace $\widetilde{\Sigma}$ of $H^2(\widetilde{K3})$ (see Appendix \ref{AppendixK3}). Up to $SO(3)$ rotations, it is defined by three vectors $\omegat_j\in H^2(\widetilde{K3})$ ($j=1,2,3$), normalized to unit length. These three 2-forms provide a complex structure and a K\"ahler form, up to $SO(3)$ rotations (we have an $\Ss^2$ of possible choices):
\begin{equation}
 \omegat = \omegat_1+i\omegat_2  \qquad\qquad \jt = (2\nut)^{1/2}\omegat_3 \:,
\end{equation}
where $\nut$ is the volume of $\widetilde{K3}$. The condition to be an elliptic fibration means that there exist two $\omegat_j$ orthogonal to $F$ and $B$. The holomorphic 2-form is a combination of them. This selects unambiguously one complex structure among the possible ones.

The metric deformations are the deformations of $\omegat_j$ that give a different 3-plane, plus the volume modulus.

\

In \cite{BHT} it is shown explicitly how to associate the complex structure deformations of the elliptically fibred $\widetilde{K3}$ with the complex structure of $T^2/\Zbb_2$, the axio-dilaton and the D7-brane positions: The vectors in $H^2(\widetilde{K3})$ orthogonal to $F$ and $B$ can be expanded in a basis of integral forms given by $\{\e_1,\alpha,\e_2,\beta,A_h,B_h,C_h,D_h\}$\label{AhBhChDh}, with $h=1,2,3,4$ (see Appendix \ref{AppendixK3} for the definition of this basis), with respect to which the metric has the block-diagonal form
\begin{equation}
\left(\begin{array}{ccc}
\begin{array}{cc} 0&2\\2&0\\ \end{array} & & \\
& \begin{array}{cc} 0&2\\2&0\\ \end{array}  & \\
& & \mathbb{D}_4^4\\
\end{array}\right)\:.
\end{equation}
$\mathbb{D}_4^4$ is the Cartan matrix of the $SO(8)^4$ group.

The basis elements $\e_1,\alpha,\e_2,\beta$ can be associated with the 2-cycles constructed by the two 1-cycles of the base $T^2/\Zbb_2$ and the two 1-cycles of the fibre $T^2$. Following the recipe given in \cite{BHT}:
\begin{equation}
 \frac{\e_1}{\sqrt{2}} = dy\wedge dy' \qquad \frac{\alpha}{\sqrt{2}} = -dx\wedge dx' \qquad \frac{\e_2}{\sqrt{2}} = dy\wedge dx' \qquad \frac{\beta}{\sqrt{2}} = dx\wedge dy'\:,
\end{equation}
where $x,y$ are coordinates on the base, while $x',y'$ are coordinates on the fibre\footnote{
We are taking a different normalization with respect to \cite{BHT}. For us $\int_{T^2/\Zbb_2}dx\,dy = 1$}.

The other sixteen cycles are the 2-cycles that shrink to zero when $\widetilde{K3}$ develops an $SO(8)^4$ singularity.

\

We are now ready to give the explicit map between the fields in the two compactifications. 
We will take an elliptically fibred $\widetilde{K3}$ that has an $SO(8)^4$ singularity. In particular we consider deformations of $\widetilde{K3}$ around the point in the moduli space defined by:
\begin{equation}\label{K3IIBpt}
 \omegat^{(o)} = \omegat^{(o)}_1+i\,\omegat^{(o)}_2 = \left( -\frac{\e_1}{2} -\frac{\alpha}{2}\right)  +i\left( \frac{\e_2}{2} + \frac{\beta}{2}\right)
	\qquad\qquad \omegat^{(o)}_3 = \frac{1}{\sqrt{2}}(B + 2 F)
\end{equation}
The 57 deformations of this point are described by the vectors $\delta\omegat_j$. These vectors are orthogonal to $\widetilde{\Sigma}=<\omegat_1^{(o)},\omegat_2^{(o)},\omegat_3^{(o)}>$ and can be expanded as:
\begin{equation}
 \delta\omegat_j= \et_i^1\left(\frac{\e_1}{2} -\frac{\alpha}{2}\right) + \et_i^2\left(\frac{\e_2}{2} -\frac{\beta}{2}\right) + \et_i^3 \frac{B}{\sqrt{2}} + \et_i^{\theta+3}\tilde{u}_{\theta+3},
\end{equation}
where $\tilde{u}_{\theta+3}$ are 16 vectors orthogonal to $<F,B,\e_1,\alpha,\e_2,\beta>$.

The explicit map for the 3d scalars is given by:
\begin{equation}\label{Map3dScalars}
\begin{array}{ccccccccccccccccc}
  IIB      &\vline& \hat{e}_i^m &\vline& \rho_2  &\vline& \hat{C}^I  &\vline& \hat{\tau},\hat{\sigma} &\vline& \hat{x}^\vartheta,\hat{y}^\vartheta  &\vline& \hat{A}_B^K  &\vline& \hat{A}_B^\vartheta &\vline& \hat{\phi},\hat{r}_B\\  &\vline&&\vline&&\vline&&\vline&&\vline&&\vline&&\vline&&\vline&\\ \hline &\vline&&\vline&&\vline&&\vline&&\vline&&\vline&&\vline&&\vline&\\
  \mbox{M-theory}  &\vline& e_i^m	  &\vline& \nu     &\vline& C_3^I     &\vline& \tilde{e}_1^1,\tilde{e}_1^2,\tilde{e}_2^1,\tilde{e}_2^2 &\vline& \tilde{e}_1^{\vartheta+3},\tilde{e}_2^{\vartheta+3} &\vline&  \tilde{e}_1^3,\tilde{e}_2^3,\tilde{e}_3^1,\tilde{e}_3^2 &\vline& \tilde{e}_3^{\vartheta+3} &\vline& \nut,\tilde{e}^3_3   \\
\end{array}\nonumber
\end{equation}

Let us explain this table. The first two columns are obvious: they are the metric moduli of $K3$ in both compactifications.

The third one is due to the point (7) at page \pageref{pt7dual}.

As it can be seen from (1), (8) and (9), the complex structure of $T^2/\Zbb_2$ and the axiodilaton go to M-theory complex structure deformations of the base and the fibre; these complex structure deformations have been identified in \cite{BHT} to the deformations of $\omegat$ in the subspace of $H^2(\widetilde{K3})$ given by $\e_1,\alpha,\e_2,\beta$, that we have called $\tilde{e}_1^1,\tilde{e}_1^1,\tilde{e}_2^1,\tilde{e}_2^2$.

The positions of the D7-branes, relative to the O7-planes, give informations on the elliptic fibration, and are associated with the complex structure deformations in the space orthogonal to $\e_1,\alpha,\e_2,\beta$ \cite{BHT}. When $\omegat$ has non-zero components along this space, some of the shrunk cycles blow up and the singularity changes. This corresponds in type~IIB to a change of the gauge group, due to some D7-branes going far from the orientifold planes.

The scalars $\hat{A}_B^K$ come from $B_2,C_2$ with one index on $\Ss^1_B$. These are mapped (see (4) at page \pageref{pt4dual}) to metric elements with one index on the fibre and one on the base. These are described by the two deformations of~$\omegat$ on the $<F,B>$ subspace, and the two deformations of $\omegat_3$ on the subspace $<\e_1,\alpha,\e_2,\beta>$. Note that they cannot be mapped to deformations of $\omegat_j$'s along cycles belonging to the $\mathbb{D}_4^4$ block: In fact a general vev for  $\hat{A}_B^K$ generates a Wilson line for the 4d~vector $A_\mu^K$ along $\Ss^1_B$; these Wilson lines do not break the gauge group on the D7-branes as they arise from $B_2$ and $C_2$. This means that these degrees of freedom cannot be mapped to deformations of $\widetilde{K3}$ that would change the $SO(8)^4$ singularity.

The scalars $\hat{A}_B^\vartheta$ go to the deformation of $\omegat_3$ along the vectors of the $\mathbb{D}_4^4$: They give the positions of the D6-branes (dual to the D7-branes) on the $T^2$ fibre. When some D7-branes are on top of each other, the fibre torus degenerates. Correspondingly the complex structure $\omegat$ is orthogonal to some 2-cycles with topology of $\Ss^2$ \cite{BHT}; this does not mean that these cycles have shrunk, because they could be not orthogonal to $\omegat_3$.\footnote{The volume of a 2-cycle on $K3$ is given by $\rho(C_2)^2=\sum_i \left| \int_{C_2} \omega_i\right|^2$.} When it happens, their sizes (given by $\omegat_3$ moduli) describe the distances between D6-branes in the degenerate fibre (the fibre degenerates in a collection of $\Ss^2$ whose size is given by $\omegat_3$ \cite{Aspinwall:1997eh}). This corresponds precisely to non-vanishing Wilson lines on the D7-branes. In this case the gauge group is broken; one can see this in type~IIB as gauge symmetry breaking due to abelian Wilson lines and in M-theory from the fact that some cycles have been blown up and the singularity has been changed.

Finally, the size of $\Ss_B^1$ and of $T^2/\Zbb_2$ go respectively to the size of the fibre and of the base of the fibration. These are given by the volume modulus of $\widetilde{K3}$ and the modulus describing the rotation on $\omegat_3$ in $F,B$ subspace ({\it i.e.} the one giving the relative size of fibre and base).

\

The map for the vectors is:
\begin{equation}\label{MapVectors}
\begin{array}{ccccccccc}
IIB &\vline& \hat{A}_\mu^K &\vline& \hat{A}_\mu^\vartheta &\vline& \hat{g}_\mu &\vline& \hat{C}_{4\mu} \\ &\vline&&\vline&&\vline&&\vline& \\ \hline &\vline&&\vline&&\vline&&\vline& \\
\mbox{M-theory} &\vline& \tilde{C}_{3\mu}^{(\e_1)},\tilde{C}_{3\mu}^{(\alpha)},\tilde{C}_{3\mu}^{(\e_2)},\tilde{C}_{3\mu}^{(\beta)} &\vline& \tilde{C}_{3\mu}^{\vartheta+3} &\vline& \tilde{C}_{3\mu}^{(F)} &\vline& \tilde{C}_{3\mu}^{(B)} \\
\end{array}\nonumber
\end{equation}

In M-theory, the vectors come all from $C_3$ along 2-cycles of $\widetilde{K3}$. In type~IIB they come from $B_2,C_2$, from the D7-branes worldvolume, from the metric and from $C_4$. Because of (3) at page \pageref{pt3dual}, the vectors coming from $B_2,C_2$ on 1-cycles of $T^2/\Zbb_2$ go to the ones coming from $C_3$ along the 2-cycles associated with these 1-cycles, {\it i.e.} $\e_1,\alpha,\e_2,\beta$ \cite{BHT}. (11) at page \pageref{pt11dual} says that the D7 $U(1)$ gauge fields go to $U(1)$ gauge fields coming from $C_3$ along the shrinking cycles (the ones giving the D7 configuration \cite{BHT}). Finally, (2) and (7) say that $g_\mu$ and $C_{4\mu}$ go respectively to $C_3$ along the fibre and $C_3$ along the base.

\

We conclude this section by a remark. The distinction between the M-theory fields corresponding to bulk and brane fields is special to the orientifold limit around which we are expanding. This means that the duality map given above works in a clean way when we are considering fluctuations of the $SO(8)^4$ vacuum, that in type~IIB corresponds to four D7-branes on top of each O7-plane, and in M-theory to $\widetilde{K3}$ having a $\mathbb{D}_4^4$ singularity. On the other hand, if we remain in the weak coupling limit region ({\it i.e.} small dilaton), then we can still trust this map \cite{Sen:1997gv}.

\subsection{F-Theory Limit and Duality in Four Dimensions}

We now derive what happens to the M-theory fields in the F-theory limit, {\it i.e.} when we take the size of the fibre to zero. To obtain this, we apply the map above to the corresponding  type~IIB limit which we have described at page \pageref{IIBlimit}. We remember that the size of the fibre is mapped under duality to the (inverse) size of $\Ss^1_B$. 

First, we note that when the fibre size vanishes, new degrees of freedom must arise to describe the dependence of the fields on the fourth dimension. They are the dual to the KK IIB modes along $\Ss_B^1$ (that in IIA are seen as string winding modes along $\Ss_A^1$).

Let us describe the behavior of the M-theory 3d fields in the F-theory limit is the following. By a field redefinition, we replace the two scalars $\nut$ and $\tilde{e}^3_3$ with the scalars related to the sizes of fibre and base $v_F,v_B$, we see that $v_F$ combine with the vector $C_{3\mu}^{(F)}$ to give the 4d~$g_{\mu\nu}$. The vector $C_{3\mu}^{(B)}$ becomes a 4d 2-form, that dualizes to a scalar. The vectors $C_{3\mu}^{(\e_1)},C_{3\mu}^{(\alpha)},C_{3\mu}^{(\e_2)},C_{3\mu}^{(\beta)}$ eat the scalars $\tilde{e}_1^3,\tilde{e}_2^3,\tilde{e}_3^1,\tilde{e}_3^2$ and become 4d vectors. Analogously,the vectors $C_{3\mu}^{\vartheta+3}$ eat the scalars $\tilde{e}_3^{\vartheta+3}$. Then, all these degrees of freedom disappear from the F-theory moduli space. In particular, the last one correspond to D7 Wilson line on $\Ss_B^1$, that disappear from the type~IIB moduli space, as they become pure gauge. In IIA/M-theory, sending the fibre to zero makes the D6-brane positions to collapse on top of each other, making irrelevant if they were separated or not before the limit; the corresponding $\Ss^2$'s shrink to zero size. In this case, only the complex structure gives the singularity type and so the gauge group after the limit. 

The 4d moduli are then given by the remaining scalars $e_i^b,C^I$ and $\nu$ form $K3$, as well as $\tilde{e}_1^1,\tilde{e}_1^2,\tilde{e}_2^1,\tilde{e}_2^2$, $\tilde{e}_1^{\vartheta+3},\tilde{e}_2^{\vartheta+3}$ and $v_B$ from $\widetilde{K3}$.
The same result has been obtained in \cite{BHLV} by considering the F-theory limit directly in M-theory.


\section{F-Theory Scalar Potential}\label{Sec:potential}

Introducing background fluxes in M-theory gives a potential for the geometric moduli. In the case of compactification on $K3\times K3$, the full scalar potential has been derived and studied in \cite{BHLV} (the problem of moduli fixing was studied previously in \cite{Lust,Kachru,BergMay,AspKall} using the Gukov-Vafa-Witten superpotential \cite{GVW}). We want to relate this potential to the type~IIB supergravity potential studied in \cite{Fer,Fer2}. To do this we have to turn on an M-theory flux that is dual to the type~IIB one. Then we take the M-theory potential generated by that flux and apply the map described so far. We will see that the result is precisely the scalar potential for gauged 4d supergravity given in \cite{Fer,Fer2}.

\subsection{M-Theory Potential}

Turning on background fluxes for $F_4=dC_3$ generates a potential for the geometric moduli of $K3\times \widetilde{K3}$, that can fix some of them. Since we want to use this background to study a 4d compactification of type~IIB, we will consider only 4-form fluxes with two legs on one $K3$ and two on the other. A flux completely on one $K3$ would be mapped to type~IIB vev's that break 4d Lorentz invariance \cite{Dasgupta:1999ss}. This flux can be expanded into a basis\footnote{We work in unit where the quantized fluxes have integral coefficients with respect to integral bases.} of $H^2(K3)\otimes H^2(\widetilde{K3})$:
\begin{equation}\label{F4flux}
 F_4 = G^{I\Lambda}\eta_I\wedge\etat_\Lambda \:.
\end{equation}
$F_4$ gives a contribution to the M2-charge. The cancellation condition for this charge is \cite{BandB}
\begin{equation}
 N_{flux}^{Mth} + N_{M2} = \frac{\chi(K3\times K3)}{24} = 24 \qquad\qquad\mbox{where }
	\qquad N_{flux}^{Mth}= \frac12 \int_{K3\times K3} F_4\wedge F_4
\end{equation}
The potential generated by the flux \eqref{F4flux} has the following expression \cite{BHLV}:
\begin{align}\label{potentialV}
  V = -\frac{2\pi}{\nu^3\nut^3} \left( \sum_i \left\Vert \Ppt[G^a\omega_i] \right\Vert^2 + \sum_j \left\Vert
    \Pp[G\,\omegat_j] \vphantom{\Ppt}\right\Vert^2\right) 
\end{align}
Let us explain the notation. The norms are relative to the metrics on $H^2(K3)$ and $H^2(\widetilde{K3})$ given by the wedge product (see \eqref{K3modmetric} in the Appendix \ref{AppendixK3}). $\Pp$ projects the vectors of $H^2(K3)$ to the subspace orthogonal to all the $\omega_i$'s. $\Ppt$ is defined analogously. $G$ and $G^a$ are two homomorphisms $G:H^2(\widetilde{K3}) \rightarrow H^2(K3)$ and
$G^a:H^2(K3)\rightarrow H^2(\widetilde{K3})$ defined as:
\begin{equation}
G \tilde{v} = (G^{I\Lambda} \Mt_{\Lambda\Sigma} \tilde{v}^\Sigma)\, \eta_I \qquad\qquad G^a v = (v^J M_{JI} G^{I\Lambda}) \etat_\Lambda \:,
\end{equation}
where $v=v^J\eta_J\,\in H^2(X)$ and $\tilde{v}=\tilde{v}^\Sigma\etat_\Sigma\, \in H^2(\Xt)$, and where $\Mt_{\Lambda\Sigma}\equiv (\etat_\Lambda \cdot \etat_\Sigma)$ and $M_{JI}\equiv (\eta_I \cdot \eta_J)$. \\$G$ and $G^a$ satisfy $(v\cdot G \tilde{v}) = (G^a v \cdot \tilde{v})$.

The potential \eqref{potentialV} is positive definite and its minima are at $V=0$. The volume moduli $\nu$ and $\nut$ are flat directions on the minima. The remaining $57+57$ moduli are encoded into $\omega_i$ and $\omegat_j$ and are generically fixed by fluxes.

\

We now want to consider the deformations around minima that correspond to type~IIB configurations. They have been studied in \cite{BHLV}. In particular, we will consider deformations of $\widetilde{K3}$ around the point given by \eqref{K3IIBpt}. 

Since we are interested in the 4d type~IIB dual scalar potential, we will consider M-theory deformations that correspond to 4d type~IIB scalars, and we will keep fixed the other ones. Moreover we will turn on $F_4$ fluxes that map to couplings of the type~IIB 4d gauged supergravity.

\subsubsection*{Map of the fluxes}

Let us see how the fluxes are transformed under the duality. In M-theory the fluxes are expectation values for $F_4=dC_3$. Because of points (3,5,11) of page \pageref{pt3dual}, when $F_4$ has two indices on one $K3$ and two on the other one, it is mapped to the bulk type~IIB fluxes $F_3,H_3$ and to D7-brane flux $F_2$ (see \cite{DenefRev} for details).

Let us focus on the 3-form IIB fluxes $F_3,H_3$. The map is given by
\begin{equation}
 F_4 = F_3\wedge dx' + H_3\wedge dy' \:,
\end{equation}
where $x',y'$ are flat coordinates on the $T^2$ fibre. If we insert \eqref{3flux} into the expression above, we get
\begin{eqnarray}\label{F4fluxMthIIB}
F_4 &=& \left\{ f_0^i\eta^i \wedge\left(-\frac{\alpha}{2}-\frac{\e_1}{2}\right) +f_2^i\eta^i \wedge\left(\frac{\alpha}{2}-\frac{\e_1}{2}\right) + f_1^i\eta^i \wedge\left(\frac{\beta}{2}+\frac{\e_2}{2}\right) +f_3^i\eta^i \wedge\left(\frac{\beta}{2}-\frac{\e_2}{2}\right) \right.\nonumber\\
 &&\left.+h_0^b\eta^b \wedge\left(-\frac{\alpha}{2}-\frac{\e_1}{2}\right)+ h_2^b\eta^b \wedge\left(\frac{\alpha}{2}-\frac{\e_1}{2}\right)+h_1^b\eta^b \wedge\left(\frac{\beta}{2}+\frac{\e_2}{2}\right)+h_3^b\eta^b \wedge\left(\frac{\beta}{2}-\frac{\e_2}{2}\right)\right\}\nonumber\\ \nonumber\\
 &=& \left\{ f_0^i\,\eta^i \wedge\omegat^{o}_1 +f_2^i\,\eta^i \wedge\tilde{u}_1 + f_1^i\,\eta^i \wedge\omegat^{o}_2 +f_3^i\,\eta^i \wedge\tilde{u}_2 \right. \\
 && \left. + h_0^b\,\eta^b \wedge\omegat^{o}_1+ h_2^b\,\eta^b \wedge\tilde{u}_1+h_1^b\,\eta^b \wedge\omegat^{o}_2+h_3^b\,\eta^b \wedge\tilde{u}_2 \right\}\:,\nonumber
\end{eqnarray}
where $\tilde{u}_1=(\frac{\e_1}{2}-\frac{\alpha}{2})$ and $\tilde{u}_2=(\frac{\e_2}{2}-\frac{\beta}{2})$ are two vectors orthogonal to $\omegat^{o}_1,\omegat^{o}_2,F,B$ and with norm~$-1$.

We see that the flux \eqref{F4fluxMthIIB} is precisely of the form \eqref{F4flux}, where the basis $\{\etat_\Lambda\}$ is given by $\{\omegat^{o}_1$, $\omegat^{o}_2$, $\omegat^{o}_3$, $\tilde{u}_1$, $\tilde{u}_2$, $\tilde{u}_3$, $\tilde{u}_{\vartheta+3}\}$; $\tilde{u}_3=\frac{B}{\sqrt{2}}$ is vector in $<F,B>$ orthogonal to $\omegat^{o}_j$, and $\tilde{u}_{\vartheta+3}$ are other 16 vectors that complete the orthonormal basis.

The formula \eqref{F4fluxMthIIB} allows us to give the precise matching between fluxes $G^{I\Lambda}$ and gauge couplings $f_K^i,h_K^i$ (this is the same result found in \cite{Lust}):
\begin{align}\label{mapFlux}
 G^{i1} = f_0^i \qquad G^{i2} = f_1^i \qquad G^{i4} = f_2^i \qquad G^{i5} = f_3^i \nonumber\\
 G^{b1} = h_0^b \qquad G^{b2} = h_1^b \qquad G^{b4} = h_2^b \qquad G^{b5} = h_3^b 
\end{align}
and all the other coefficients are zero.

\

The M2-charge carried by the flux \eqref{F4fluxMthIIB} is equal to the type~IIB  D3-charge \eqref{F3H3charge}, {\it i.e.} $N_{flux}^{Mth} = N_{flux}^{OR}$.

\vskip 7mm

\subsection{M-Theory Potential around the \boldmath$SO(8)^4$ Point}

To get the potential \eqref{potentialV} as a function of the moduli, we have to take a suitable expansion of the vectors that define the 3-planes $\Sigma$ and $\widetilde{\Sigma}$:
\begin{equation}\label{omegaexp}
\omega_i = a_i^p \omega_p^{o} + e_i^b u_b \qquad\qquad 
\omegat_j = \tilde{a}_i^q \omegat_q^{o} + \tilde{e}_j^c \tilde{u}_c \:.
\end{equation}
The basis $\{\etat_\Lambda\}$ has been defined before. We take the orthonormal basis $\{\eta_I\}$ to be also made up of the 3 positive norm vectors $\omega_i^o$ and 19 negative norm vectors $u_b$. The vectors $\omega_i^o$ define a 3-plane $\Sigma$ in $H^2(K3)$, that is the point in the $K3$ moduli space around which we are expanding.

The coefficients $a_i^p$ and $\tilde{a}_j^q$ depend respectively on $e_i^b$ and $\tilde{e}_j^c$ once we require $\omega_i$ and $\omegat_j$ to satisfy $\omega_i\cdot \omega_n = \delta_{in}$ and $\omegat_j\cdot \omegat_m = \delta_{jm}$, that means
\begin{equation}\label{Constrae}
\sum_pa_i^pa_n^p = \delta_{in} + \sum_be_i^be_n^b \qquad\mbox{and}\qquad \sum_q\at_j^q\at_m^q = \delta_{jm} + \sum_c\et_j^c\et_m^c\:.
\end{equation}
We fix the arbitrariness due to $SO(3)$ rotation in $\Sigma$ and $\widetilde{\Sigma}$ by requiring 
\begin{equation}\label{Constraebis}
 a_1^2 = a_2^1 \:, \qquad a_3^1 = a_3^2 = 0 \:, \qquad\qquad \tilde{a}_1^2 = \tilde{a}_2^1 \:, \qquad \tilde{a}_3^1 = \tilde{a}_3^2 = 0 \:.
\end{equation}
First, we need to compute $\Pp[G\omegat_j]$ and $\Ppt[G^a\omega_i]$:
\begin{equation}\label{PpPpt}
 \Pp[G\omegat_j] = G\omegat_j -\sum_i (\omega_i \cdot G\omegat_j) \,\omega_i \qquad
 \Ppt[G^a\omega_i] = G^a\omega_i -\sum_j (\omegat_j \cdot G^a\omegat_i) \,\omegat_j \:.
\end{equation}
Inserting the expansions \eqref{omegaexp} into \eqref{PpPpt}, we get:
\begin{eqnarray}
\Pp[G\,\omegat_j] &=& \left[\tilde{a}_j^q {G^i}_{q}+\tilde{e}_j^c {G^i}_{c} - \sum_n Q_{nj} a_n^i\right] \omega_i^{o} +
		\left[\tilde{a}_j^k {G^b}_{k}+\tilde{e}_j^c {G^b}_{c} - \sum_\ell Q_{\ell j} e_\ell^b\right] u_b\nonumber\\
\Ppt[G^a\omega_i] &=& \left[a_i^p {G_p}^j+e_i^b {G_b}^j - \sum_m\tilde{Q}_{mi} \tilde{a}_m^j\right] \omegat_j^{o} +
		\left[a_i^h {G_h}^c+e_i^b {G_b}^c - \sum_s\tilde{Q}_{si} \tilde{e}_s^c\right] \tilde{u}_c \:,
\end{eqnarray}
where 
\begin{eqnarray}
Q_{nj} &=& a_n^p\tilde{a}_j^qG_{pq}+a_n^p \tilde{e}_j^c G_{pc} + e_n^b \tilde{a}_j^q G_{bq}+e_n^b\tilde{e}_j^c G_{bc}\nonumber\\
\tilde{Q}_{mi} &=& \tilde{a}_m^q a_i^pG_{pq}+\tilde{a}_m^q e_i^b G_{bq} + \tilde{e}_m^c a_i^p G_{pb}+\tilde{e}_m^c e_i^b G_{bc} \:.\nonumber
\end{eqnarray}
The potential is then given by
\begin{eqnarray}\label{potentialV2}
  V &=& -\frac{2\pi}{\nu^3\nut^3} \left\{\sum_{i,j} \left[\tilde{a}_j^q {G^i}_{q}+\tilde{e}_j^c {G^i}_{c} - \sum_n Q_{nj} a_n^i\right]^2 - \sum_{j,b}\left[\tilde{a}_j^k {G^b}_{k}+\tilde{e}_j^c {G^b}_{c} - \sum_\ell Q_{\ell j} e_\ell^b\right]^2 \right.\nonumber\\
	&& +\left.\sum_{i,j} \left[a_i^p {G_p}^j+e_i^b {G_b}^j - \sum_m\tilde{Q}_{mi} \tilde{a}_m^j\right]^2 
	- \sum_{i,c}\left[a_i^h {G_h}^c+e_i^b {G_b}^c - \sum_s\tilde{Q}_{si} \tilde{e}_s^c\right]^2\right\}\:.
\end{eqnarray}
If we consider the case where $G^{ic}=G^{bj}=0$, $G^{ij}=\frac{g_{i-1}}{\sqrt{2}}\delta^{ij}$ and $G^{bc}=\frac{\ell_b}{\sqrt{2}} \delta^{bc}$, then it takes the simplified form
\begin{eqnarray}\label{potVsimpl}
 V &=& -\frac{\pi}{\nu^3\nut^3} \left\{\sum_{i,j}g_{i-1}^2 \left[\left(\tilde{a}_j^i\right)^2 + 
	 \left(a_j^i\right)^2\right] - \sum_{j,b} \ell_b^2\left[  \left(\tilde{e}_j^b\right)^2 
		+ \left(e_j^b\right)^2\right]  -2\sum_{jm} \hat{Q}_{jm}^2  \right\} \:,
\end{eqnarray}
where 
\begin{equation}
 \hat{Q}_{jm} = \sum_i g_{i-1}\tilde{a}_j^i a_m^i + \sum_b \ell_b \tilde{e}_j^b e_m^b \:.
\end{equation}

\

We want to compare this potential with the type~IIB one \cite{Fer,Fer2}. Using the map \eqref{mapFlux}, we write the coefficients of the expansion \eqref{F4fluxMthIIB} for the 4-form flux in terms of $g_{i-1}$ and $\ell_\beta$:
\begin{equation}
 f_0^1=g_0 \qquad f_1^2=g_1 \qquad h_2^1=\ell_1 \qquad h_3^2=\ell_2
\end{equation}
all the other coefficients vanish. 

We start from the flux that in type~IIB leads to \eqref{potIIB1}, {\it i.e.} we take $\ell_1=\ell_2=0$. To get the potential in terms of the moduli, we write $a_i^p$ and $\tilde{a}_j^q$ explicitely in terms of\footnote{Their expressions are given in Appendix \ref{Appendixae}} $e_i^b$ and $\et_j^c$. The result is
\begin{eqnarray}\label{potVprel}
 V &=& \frac{\pi}{\nu^3\nut^3} \left\{g_0^2 \left(e_1^2 + \tilde{e}_1^2 +2 e_1^2  \tilde{e}_1^2\right) + 
	g_1^2 \left(e_2^2 + \tilde{e}_2^2 +2 e_2^2  \tilde{e}_2^2\right) + 4 g_0 g_1 (e_1\cdot e_2)(\tilde{e}_1\cdot \tilde{e}_2)    \right\}\nonumber\\
   &=& \frac{\pi}{\nu^3\nut^3} \left\{g_0^2 \left(e_1^2 + \tilde{e}_1^2 \right) + g_1^2 \left(e_2^2 + \tilde{e}_2^2 \right) 
	+ 2 \sum_{b,c} \left( g_0 e_1^b \et_1^c + g_1 e_2^b \et_2^c \right)^2  \right\} \:,
\end{eqnarray}
where $e_i^2=\sum_b e_i^b e_i^b$ and $(e_1\cdot e_2)=\sum_b e_1^b e_2^b$, and the same for the tilded quantities.

\

From the analysis in \cite{BHLV}, we know that this flux potential must fix\footnote{The condition for $(\omega_i,\omega_j)$ to be a minimum of the potential is that the flux homomorphism $G$ maps $\omegat_j$ to $\omega_i$ and {\it viceversa}. With the choice of flux we made (diagonal in the bases $\omega_i^{(o)},u_b$ and $\omegat_j^{(o)},\tilde{u}_c$) the condition is obviously satisfied for $\omega_i=\omega_i^{(o)}$ and $\omegat_j=\omegat_j^{(o)}$.
} $\omega_1$ and $\omega_2$ to $\omega_1^o$ and $\omega_2^o$, as well as $\omegat_1$ and $\omegat_2$ to $\omegat_1^o$ and $\omegat_2^o$. This is manifest from the form \eqref{potVprel}: $V$ is positive definite and its minimus is at $e_i^b$=0 ($i=1,2$) and $\et_j^c$=0 ($j=1,2$).

The condition $V=0$ also fixes $\et_j^3$ ($j=1,2$). Then fluxes generate a mass term for them. In type~IIB these two scalars are related (see table at page \pageref{Map3dScalars}) to the fourth component of two 4d vector fields. This suggests that, because of 4d Lorentz invariance in type IIB, that the corresponding 4d vectors acquire a mass. 
In fact, in \cite{BHLV} it has been shown that the associated M-theory 3d gauge field acquire the same mass as the scalar. When they combine together in the F-theory limit, the resulting 4d vector is massive. This mass has also been found in \cite{Fer} by studying IIB with flux directly. Hence we see that the M-theory potential also gives informations on type IIB vectors.

From \cite{BHLV} we also know that if $g_0=g_1$ then the minima are $\Nn=1$ supersymmetric in 4d, that is precisely what happens in the type~IIB dual (see (2) at page \pageref{caso2}).

\

The last step to find the type~IIB expression of \cite{Fer}, is to write the $\et_i^b$ moduli in terms of the type~IIB moduli at the $SO(8)^4$ point ({\it i.e.} 4 D7 on top of each orientifold). We already have the prescription. The type~IIB deformations are encoded into the following expansion of $\omegat$ \cite{BHT}:
\begin{eqnarray}\label{omegatSO8}
 \omegat &=& \omegat_1+i\omegat_2 = \frac{1}{\sqrt{\tau_2\sigma_2}}\left\{ -\frac{\alpha}{2}  + \tau \,\frac{\e_2}{2} + \sigma\,\frac{\beta}{2}+ \tau\,\sigma \,\frac{\e_1}{2}  \right\}\nonumber\\
	&=& \frac{1}{2\sqrt{\tau_2\sigma_2}}\left\{\left[(-1+\sigma_1\tau_1-\sigma_2\tau_2)\omegat_1^o + (\sigma_1+\tau_1)\omegat_2^o + (-1-\sigma_1\tau_1+\sigma_2\tau_2)\tilde{u}_1 + (\sigma_1 - \tau_1)\tilde{u}_2  \right]\right.\nonumber\\
	&& \left. + i \left[(\sigma_1\tau_2+\sigma_2\tau_1)\omegat_1^o + (\sigma_2+\tau_2)\omegat_2^o + (-\sigma_1\tau_2-\sigma_2\tau_1)\tilde{u}_1 + (\sigma_2 - \tau_2)\tilde{u}_2\right]\right\}\:.
\end{eqnarray}
$\omegat_3$ is taken to live in $F,B$ subspace, as we are interested in the 4d result: The orthogonal deformations describe Wilson lines for the type~IIB gauge fields that go to zero after the F-theory limit \cite{BHLV}. The modulus controlling the direction in $F,B$ ($\et_3^3$) is the one that is used to take the F-theory limit. As it is explained in \cite{BHLV} it also goes away from the F-theory moduli space. This can also be seen by noting that it is mapped to one component of the 4d metric fluctuations, that we want to keep massless.

With the prescription \eqref{omegatSO8}, we derive the expression for $\at_1^q$, $\at_2^q$ and $\et_1^b$, $\et_2^b$, while we keep the $\at_3^q$ and $\et_3^q$ as in the Appendix \ref{Appendixae}:
\begin{eqnarray}
 &&\at_1^1=-\frac{1}{\sqrt{\tau_2\sigma_2}}\, \mbox{Re}X^0 \qquad \at_1^2=-\frac{1}{\sqrt{\tau_2\sigma_2}}\, \mbox{Re}X^1 \qquad \at_2^1=-\frac{1}{\sqrt{\tau_2\sigma_2}}\, \mbox{Im}X^0 \qquad \at_2^2=-\frac{1}{\sqrt{\tau_2\sigma_2}}\, \mbox{Im}X^1\nonumber\\
 &&\et_1^1=\frac{1}{\sqrt{\tau_2\sigma_2}}\, \mbox{Re}X^2 \qquad \et_1^2=-\frac{1}{\sqrt{\tau_2\sigma_2}}\, \mbox{Re}X^3 \qquad \et_2^1=\frac{1}{\sqrt{\tau_2\sigma_2}}\, \mbox{Im}X^2 \qquad \et_2^2=-\frac{1}{\sqrt{\tau_2\sigma_2}}\, \mbox{Im}X^3\:,\nonumber
\end{eqnarray}
where $X^0,X^1,X^2$ and $X^3$ are the function of $\tau$ and $\sigma$ given in \eqref{X0123}. We note that at $\tau=\sigma=i$, the $\et_i^b$ vanish.

We insert these expressions in the potential above, and we find
\begin{eqnarray}
 V &=& \frac{\pi}{\nu^3\nut^3} \left\{\frac{2}{\tau_2\sigma_2}\left[g_0^2 |X^0|^2 (1+ 2e_1^2) + g_1^2 |X^1|^2 (1+ 2e_2^2) + \right.\right.\nonumber\\
 && \left.\left. + 2 g_0 g_1 (e_1\cdot e_2) (X^0\bar{X}^1+\bar{X}^0X^1)\right] -2 \left[ g_0^2  (1+ e_1^2) + g_1^2 (1+ e_2^2)\right] \right\}\:.
\end{eqnarray}
If we now consider the expressions \eqref{KpotIIB} for $\widetilde{\mathcal{K}}$ and $\widehat{\mathcal{K}}$, the fact that $\nu=\rho_2$, and a rescaling of the potential due to the F-theory limit\footnote{$\nut\sim v_B v_F - v_F^2$ is mapped to $v_B/R_B^2$. When doing the limit $v_F\rightarrow 0$ and scaling the coordinate and the metric to get a 4d finite result, the $R_B^2$ factor disappear form $V$ ($v_B=e^{-2\phi}$).}, we arrive precisely to the potential \eqref{potIIB1}.

\

When we turn on non vanishing $\ell_1$ and $\ell_2$, the analysis of \cite{BHLV} tells that not only $e_1,e_2$ and $\et_1,\et_2$ are fixed, but also $e_3^1,e_3^2$ and $\et_3^1,\et_3^2$. Applying the duality map, this is translated to saying that the moduli $e_3^1,e_3^2$ are fixed and that all the four vector fields get a mass. This is precisely the result of \cite{Fer,Fer2}, reported at page \pageref{ell1ell2IIBresults}.

We are also able to derive the type~IIB formula \eqref{potentialIIBallIsom}. We consider the formula \eqref{potVsimpl} for the potential, and let all the $\ell_b$ to be different from zero, except for $b=3$ (that corresponds to have a 4-form flux along $F$ or $B$, which would break Lorentz invariance). We define $\ell_\beta$ with $\beta=1,...18$ such that $\ell_\beta=\ell_b$ for $b=1,2$ and $\ell_\beta=\ell_{b-1}$ for $b=4,...,19$.

To match with the formula \eqref{potentialIIBallIsom}, we put all $e_i^b$ to zero and we let $\et_3^b$ vanish.

Then, we have to insert the expressions for $\at_j^q$ and $\et_j^c$ in terms of the type~IIB fields. When we allow for D7-brane movement ({\it i.e.} we introduce the D7 moduli), the general form for $\omegat$ is \cite{BHT}
\begin{eqnarray}\label{omegatSO8plusD7def}
 \omegat &=& \omegat_1+i\omegat_2 = \frac{1}{\sqrt{\left[\tau_2\sigma_2-y^2/2\right]}}\left\{ -\frac{\alpha}{2}  + \tau \,\frac{\e_2}{2} + \sigma\,\frac{\beta}{2}+ \left(\tau\,\sigma -\frac{z^2}{2}\right) \,\frac{\e_1}{2} +\frac{z^\vartheta}{\sqrt{2}}\,\tilde{u}_{\vartheta+3} \right\} \:,
\end{eqnarray}
where $z^2=\sum_\vartheta z^\vartheta z^\vartheta$, as well as $y^2 =\sum_\vartheta y^\vartheta y^\vartheta$.
The vectors $\tilde{u}_{\vartheta+2}$ ($\vartheta=1,...,16$) form an orthonormal basis in the subspace of $H^2(\widetilde{K3})$ generated by $A_h,B_h,C_h,D_h$ introduced at page \pageref{AhBhChDh}. In \cite{BHT}, $z^\vartheta=x^\vartheta+i\,y^\vartheta$ have been identified with the positions of the D7-branes on $T^2/\Zbb_2$. Indeed, when they are all zero, $\omegat$ is orthogonal to all $A_h,B_h,C_h,D_h$ and $\widetilde{K3}$ develops an $SO(8)^4$ singularity. This corresponds to 4 D7-branes on top of each O7-plane in type~IIB language.

From \eqref{omegatSO8plusD7def} we can derive the expansions of $\omegat_1$ and $\omegat_2$:
\begin{eqnarray}
\omegat_1 &=& \frac{1}{2\sqrt{\tau_2\sigma_2-y^2/2}}\left\{(-1+\sigma_1\tau_1-\sigma_2\tau_2-\mbox{Re}[z^2/2])\,\omegat_1^o + (\sigma_1+\tau_1)\,\omegat_2^o \right.\nonumber\\ &&\left. + (-1-\sigma_1\tau_1+\sigma_2\tau_2+\mbox{Re}[z^2/2])\,\tilde{u}_1 + (\sigma_1 - \tau_1)\tilde{u}_2  +2 \frac{\mbox{Re}[z^\vartheta]}{\sqrt{2}}\,\tilde{u}_{\vartheta+3} \right\}\\
\omegat_2 &=& \frac{1}{2\sqrt{\tau_2\sigma_2-y^2/2}}\left\{(\sigma_1\tau_2+\sigma_2\tau_1-\mbox{Im}[z^2/2])\,\omegat_1^o + (\sigma_2+\tau_2)\,\omegat_2^o \right.\nonumber\\ &&\left. + (-\sigma_1\tau_2-\sigma_2\tau_1+\mbox{Im}[z^2/2])\,\tilde{u}_1 + (\sigma_2 - \tau_2)\tilde{u}_2  +2 \frac{\mbox{Im}[z^\vartheta]}{\sqrt{2}}\,\tilde{u}_{\vartheta+3} \right\} \:.
\end{eqnarray}

Analogously as before, we get
\begin{eqnarray}
 &&\at_1^1=-\frac{1}{\sqrt{\tau_2\sigma_2-y^2/2}}\, \mbox{Re}X^0 \qquad \at_1^2=-\frac{1}{\sqrt{\tau_2\sigma_2-y^2/2}}\, \mbox{Re}X^1 \qquad \at_2^1=-\frac{1}{\sqrt{\tau_2\sigma_2-y^2/2}}\, \mbox{Im}X^0 \nonumber\\ && \at_2^2=-\frac{1}{\sqrt{\tau_2\sigma_2-y^2/2}}\, \mbox{Im}X^1 \qquad
 \et_1^1=\frac{1}{\sqrt{\tau_2\sigma_2-y^2/2}}\, \mbox{Re}X^2 \qquad \et_1^2=-\frac{1}{\sqrt{\tau_2\sigma_2-y^2/2}}\, \mbox{Re}X^3 \nonumber\\ && \et_2^1=\frac{1}{\sqrt{\tau_2\sigma_2-y^2/2}}\, \mbox{Im}X^2 \qquad\qquad \et_2^2=-\frac{1}{\sqrt{\tau_2\sigma_2-y^2/2}}\, \mbox{Im}X^3 \nonumber\\ && \et_1^{\vartheta+3}=\frac{1}{\sqrt{\tau_2\sigma_2-y^2/2}}\,\mbox{Re}[X^{\vartheta+3}] \qquad \et_2^{\vartheta+3}=\frac{1}{\sqrt{\tau_2\sigma_2-y^2/2}}\,\mbox{Im}[X^{\vartheta+3}] \:,\nonumber
\end{eqnarray}
where $X^0,X^1,X^2,X^3$ are now given by \eqref{X0123bis}, and $X^{\vartheta+3}=\frac{z^\vartheta}{\sqrt{2}}$.

Putting these expressions in \eqref{potVsimpl} and taking $e_i^b=0$ and $\et_3^c=0$ as said above, we get 
\begin{eqnarray}\label{finalV}
 V &=& \frac{\pi}{\nu^3\nut^3} \left\{\frac{2}{\sqrt{\tau_2\sigma_2}}\left[g_0^2 |X^0|^2 + g_1^2 |X^1|^2 +\sum_{\beta=1}^{18} \ell_\beta |X^{\beta+1}|^2  + \right] -2 \left[ g_0^2  + g_1^2 \right] \right\}.
\end{eqnarray}
This is the same expression as \eqref{potentialIIBallIsom}, once we take into account the F-theory limit as before. The coefficients $\ell_{\beta}$ with $\beta=3,...,18$ can be associated with type~IIB $F_2$ flux on the D7 worldvolume \cite{Fer2,DenefRev}. They are related to M-theory $F_4$ flux in the directions of cycles controlling the positions of the branes. When they are switched on, they stabilize all $\et_j^c$ ($c\not=3$), as one can see from the M-theory analysis \cite{BHLV}. On the other hand, from type~IIB point of view we have gauged all the isometries with all the gauge fields. In both cases we conclude that the corresponding gauge fields have taken a non-zero mass.

\

In conclusion, we have seen how to derive the 4d type~IIB flux scalar potential using the 3d M-theory dual one. We have presented in detail the case of type~IIB on $K3\times T^2/\Zbb_2$ and M-theory on $K3\times K3$. We have reported the derivation of the type~IIB potential putting some moduli to zero (for example \eqref{finalV} is written after taking $e_i^b=0$), but we could write down the scalar potential (from M-theory) by keeping all the type~IIB moduli arbitrary. In this way we would get a more complete form for the type~IIB potential (we have not reported this expression here, because we wanted to show the match between expressions already derived in different theories). So we have seen that, in order to have the type~IIB potential for all the moduli, one simple way is to take the M-theory one and apply the duality map described in this paper.

We have also seen that the M-theory potential give informations on type IIB vectors. This happens because the potential depends on the 3d scalars that in type IIB become 4d vector degrees of freedom. When the corresponding M-theory moduli are stabilized, the scalars get a mass in M-theory, that signals a mass for the corresponding type IIB vectors.



\vskip 2 cm
\begin{center} \textbf{Acknowledgements} \end{center}

I would like to thank Arthur Hebecker for stimulating discussions and valuable suggestions, Andreas Braun and Christoph L\"udeling for useful discussions. Particular thanks go to Frederik Denef for reading the draft and giving enlightened comments. This work was supported by SFB-Transregio 33  "The Dark Universe" by Deutsche Forschungsgemeinschaft (DFG).

\newpage


\appendix

\section{Duality with Heterotic \boldmath$E_8\times E_8$}\label{DualWithHet}

The map we have studied in section \ref{ExplicitMap} can be extended to Heterotic compactification on $K3\times T^3$. In fact M-theory on $K3$ is dual to Heterotic $E_8\times E_8$ on $T^3$ \cite{Witten:1995ex}. In this duality, the M-theory geometric moduli of $K3$ are mapped to Heterotic geometric moduli of $T^3$, to the dilaton, to the axions coming from the 2-form $B$ and to Wilson lines on the three 1-cycles of $T^3$. The vectors coming from $C_3$ are mapped to the Heterotic vectors plus vectors coming from the metric and $B$ on $T^3$. 

But Heterotic $E_8\times E_8$ on $T^2$ is also dual to F-theory on $K3$ \cite{LustDuality}. The duality is seen by taking the decompactification limit of one $\Ss^1$ of $T^3$ in the duality above. This corresponds precisely to the F-theory limit. In particular, one can see that type~IIB on $T^2/\Zbb\times \Ss^1_B$ is dual to Heterotic on $T^2 \times \Ss^1_B$. Taking the limit $R_B\rightarrow \infty$ in both cases leads to the duality between F-theory on $K3$ and Heterotic on $T^2$.

\

Let us see what are the fields coming from compactification of Heterotic theory on $K3\times T^2\times \Ss^1_B$.\\
We start by considering what are the 3d $U(1)$ vectors (that we will denote by a hat):
\begin{description}
\item 2 vector $\hat{g}^{B}_\mu$ and $\hat{B}^B_\mu$ from the metric $g$ and the 2-form $B$ with one index on $\Ss^1_B$.
\item 4 vectors $\hat{g}^c_\mu$ and $\hat{B}^c_\mu$ ($c=1,2$) from the metric $g$ and the 2-form $B$ with one index on $T^2$.
\item 16 vectors $\hat{\mathcal{A}}_\mu^\vartheta$ ($\vartheta=1,...,16$) from the 16 $U(1)$ Heterotic gauge fields.
\end{description}
The (real) scalars (again, we will denote the 3d spectrum with a hat) are:
\begin{description}
\item 58 scalars $\hat{\nu},\hat{e}_i^b$ from the metric on $K3$.
\item 22 scalars $\hat{B}^I$ $I=1,...,22$ from $B$ on $K3$.
\item 3 scalars $\hat{v}_{T^2},\hat{t}$ from the metric on $T^2$ (respectively volume and complex structure).
\item 1 scalar $\hat{\varphi}$ from the dilaton.
\item 1 scalar $\hat{b}$ from $B$ on $T^2$.
\item 1 scalar $\hat{r}^h_B$ from the metric on $\Ss^1_B$.
\item 4 scalars $\hat{g}^c_B$ and $\hat{B}^c_B$ ($c=1,2$) from $g,B$ with one index on $T^2$ and one on $\Ss^1_B$.
\item 32 scalars $\hat{\mathcal{A}}_c^\vartheta$ ($\vartheta=0,...,16$ and $c=1,2$) from Wilson lines along $T^2$.
\item 16 scalars $\hat{\mathcal{A}}_B^\vartheta$ ($\vartheta=0,...,16$) from Wilson lines along $\Ss^1_B$.
\end{description}

Summarizing, we have 22 vectors and 138 scalars.

\

We can now complete the tables \eqref{Map3dScalars} and \eqref{MapVectorsHet} with the row corresponding to the Heterotic theory on $K3\times T^2\times \Ss^1_B$. The map between the scalars is:
\begin{equation}\label{Map3dScalarsHet}
\begin{array}{ccccccccccccccccc}
  IIB      &\vline& \hat{e}_i^m &\vline& \rho_2  &\vline& \hat{C}^I  &\vline& \hat{\sigma},\hat{\tau},\hat{\phi} &\vline& \hat{x}^\vartheta,\hat{y}^\vartheta  &\vline& \hat{A}_B^K  &\vline& \hat{A}_B^\vartheta &\vline& \hat{r}_B\\  &\vline&&\vline&&\vline&&\vline&&\vline&&\vline&&\vline&&\vline&\\ \hline &\vline&&\vline&&\vline&&\vline&&\vline&&\vline&&\vline&&\vline&\\
  \mbox{M-theory}  &\vline& e_i^m	  &\vline& \nu     &\vline& C_3^I     &\vline& \tilde{e}_1^1,\tilde{e}_1^2,\tilde{e}_2^1,\tilde{e}_2^2,\tilde{v}_B &\vline& \tilde{e}_1^{\vartheta+3},\tilde{e}_2^{\vartheta+3} &\vline&  \tilde{e}_1^3,\tilde{e}_2^3,\tilde{e}_3^1,\tilde{e}_3^2 &\vline& \tilde{e}_3^{\vartheta+3} &\vline& \tilde{v}_F  \\ &\vline&&\vline&&\vline&&\vline&&\vline&&\vline&&\vline&&\vline&\\ \hline &\vline&&\vline&&\vline&&\vline&&\vline&&\vline&&\vline&&\vline&\\
  \mbox{Heterotic}  &\vline& \hat{e}_i^m	  &\vline& \hat{\nu}     &\vline& \hat{B}^I     &\vline& \hat{t},\hat{b},\hat{\varphi},\hat{v}_{T^2} &\vline& \hat{\mathcal{A}}_c^\vartheta &\vline&  \hat{g}^c_B,\hat{B}^c_B &\vline& \hat{\mathcal{A}}_B^\vartheta &\vline& \hat{r}^h_B   \\
\end{array}\nonumber
\end{equation}

\

The table \eqref{MapVectors} for the map between vectors becomes:
\begin{equation}\label{MapVectorsHet}
\begin{array}{ccccccccc}
IIB &\vline& \hat{A}_\mu^K &\vline& \hat{A}_\mu^\vartheta &\vline& \hat{g}_\mu &\vline& \hat{C}_{4\mu} \\ &\vline&&\vline&&\vline&&\vline& \\ \hline &\vline&&\vline&&\vline&&\vline& \\
\mbox{M-theory} &\vline& C_{3\mu}^{(\e_1)},C_{3\mu}^{(\alpha)},C_{3\mu}^{(\e_2)},C_{3\mu}^{(\beta)} &\vline& C_{3\mu}^{\vartheta+3} &\vline& C_{3\mu}^{(F)} &\vline& C_{3\mu}^{(B)} \\&\vline&&\vline&&\vline&&\vline& \\ \hline &\vline&&\vline&&\vline&&\vline& \\
\mbox{Heterotic} &\vline& \hat{g}^c_\mu,\hat{B}^c_\mu &\vline& \hat{\mathcal{A}}_\mu^\vartheta &\vline& \hat{g}^{B}_\mu &\vline& \hat{B}^B_\mu \\
\end{array}\nonumber
\end{equation}

\

The F-theory limit corresponds to decompactifying $\Ss^1_B$ on the Heterotic side. This limit is trivial as the limit in type~IIB. One immediately see that the scalars $\hat{g}^c_B,\hat{B}^c_B$ and $\hat{\mathcal{A}}_B^\vartheta$ become the fourth component of the vectors $\hat{g}^c_\mu,\hat{B}^c_\mu$ and $\hat{\mathcal{A}}_\mu^\vartheta$, that become 4d vectors. The scalar $r^h_B$ combines with the vector $\hat{g}^{B}_\mu$ to give the fluctuations of the 4d metric. Finally, the vector $\hat{g}^{B}_\mu$ becomes a 4d 2-form, that is dual to a scalar. One recovers the spectrum of Heterotic on $K3\times T^2$.

\section{Some Facts on \boldmath$K3$}\label{AppendixK3}

Here, we will only collect the facts that we need; for 
a comprehensive review on $K3$ see e.g. \cite{a96}.
The Hodge numbers of $K3$ are well known: $h^{2,0}=1$ and $h^{1,1}=20$.

The metric moduli space of $K3$ is 58-dimensional. These 58 moduli can be organized as one modulus $\nu$ giving the volume of $K3$ and 57 moduli coming from the hyper-K\"ahler structure. In fact, the metric on $K3$ is fixed (up to
the overall factor, {\it i.e.} the volume) once one gives a three-dimensional plane in $H^2(K3)$. This plane is spanned by the three vectors $\omega_i$ ($i=1,2,3$) that give the $SU(2)$
structure of $K3$. They satisfy the conditions: 
\begin{align}
  \int_{K3} \omega_i\wedge\omega_j = \delta_{ij} \:.
\end{align}
From the $\omega_i$'s and the volume $\nu$, we can construct the K\"ahler and holomorphic two-form,
\begin{align}
  \omega&=\omega_1 +i \omega_2\,, &j&= \sqrt{2 \nu}\,\omega_3\,.
\end{align}
The metric is invariant under $SO(3)$ rotations of the $\omega_i$. Note that we have throughout this
work used the same letters for two-forms, their associated cohomology classes and the Poincar\'e-dual
cycles.

The 57 moduli are associated with the deformations of the three-plane
$\Sigma=\left<\omega_1,\omega_2,\omega_3\right>$ inside the space of 2-forms $H^2(K3)$. They are given by the deformations $\delta \omega_i$'s of the $\omega_i$'s in the space $R_\Sigma$ orthogonal to $\Sigma$. Orthogonality is defined using the natural metric:
\begin{align}\label{K3modmetric}
  (v \cdot w ) \equiv \int_{K3} v \wedge w \qquad \forall\; v,w\in H^2(K3) 
\end{align}
This metric has signature $(3,19)$. On $R_\Sigma$ it is negative definite.

\


Any vector in the lattice $H_2(K3,\mathbb{Z})$ of integral cycles of an elliptically fibred $K3$ can be written as
\begin{align}
  D=p^{i}e^{i}+p_{j}e_{j}+q_{I}E_{I},\label{H2}
\end{align}
where $i,j$ run from zero to three and $I,J$ from $1$ to $16$. The $p_{i}$ as well as the $p^{i}$ are
all integers. The $q_{I}$ must fulfill the conditions $\sum_{I=1..8}q_{I}=2\mathbb{Z}$ ($\sum_{I=9..16}q_{I}=2\mathbb{Z}$) and that $\forall I=1,...,8$ ($\forall I=9,...,16$) the $q_{I}$ are \emph{all} integer or \emph{all}
half-integer. The only nonvanishing inner products among the vectors of the basis of $H_2(K3)$ used in \eqref{H2} are
\begin{align}
  E_{I}\cdot E_{J}=-\delta_{IJ} \hspace{1cm} e^{i}\cdot e_{j}=\delta^{i}_{j} \ .
\end{align}
The cycles which have vanishing periods at the $SO(8)^4$ point are given by \cite{BHT}:
\begin{align}\label{AtoD}
  \text{
    \begin{tabular}[h]{c|c|c|c|c}
      & $A$&$B$&$C$&$D$\\
      \hline
      \rule[12pt]{0pt}{1pt}
      1&$E_{7}-E_{8}$&$-E_{15}+E_{16}$&
      $-e_2-E_{1}+E_{2}$&$e_2+E_{9}-E_{10}$\\
      2&$E_{6}-E_{7}$&$-E_{14}+E_{15}$ &
      $-E_{2}+E_3$&$E_{10}-E_{11}$\\
      3&$-e_1-E_{5}-E_{6}$&$e_1+E_{13}+E_{14}$&
      $-E_{3}+E_{4}$&$E_{11}-E_{12}$\\
      4&$E_{5}-E_{6}$&$-E_{13}+E_{14}$&
      $-E_{3}-E_{4}$&$E_{11}+E_{12}$.
    \end{tabular}}
\end{align}
One can check that their intersection matrix is given by the Cartan matrix of $D_4^4$.

\section{Expression of \boldmath$a_i^p$ in terms of \boldmath$e_i^b$}\label{Appendixae}

 Taking into account the constraints \eqref{Constrae} and \eqref{Constraebis}, we find the expressions of $a_i^p$ in terms of $e_i^b$:
\begin{eqnarray}
 &&a_3^3 = \sqrt{1+e_3^2} \qquad a_3^1=a_3^2=0 \qquad a_1^3=a_2^3=0\nonumber\\
 &&a_1^1= \frac{1+e_1^2+\sqrt{1+e_1^2+e_2^2+e_1^2e_2^2-(e_1\cdot e_2)^2}}{\sqrt{2+e_1^2+e_2^2+2\sqrt{1+e_1^2+e_2^2+e_1^2e_2^2-(e_1\cdot e_2)^2}}}\nonumber\\
 &&a_2^2= \frac{1+e_2^2+\sqrt{1+e_1^2+e_2^2+e_1^2e_2^2-(e_1\cdot e_2)^2}}{\sqrt{2+e_1^2+e_2^2+2\sqrt{1+e_1^2+e_2^2+e_1^2e_2^2-(e_1\cdot e_2)^2}}}\nonumber\\
 &&a_1^2 = a_2^1 = \frac{(e_1\cdot e_2)}{\sqrt{2+e_1^2+e_2^2+2\sqrt{1+e_1^2+e_2^2+e_1^2e_2^2-(e_1\cdot e_2)^2}}}\nonumber
\end{eqnarray}
where $e_i^2=\sum_b e_i^b e_i^b$ and $(e_1\cdot e_2)=\sum_b e_1^b e_2^b$.

The expressions for $\tilde{a}_j^q$ are obtained substituting $e_i^b$ with $\tilde{e}_j^c$.


\end{document}